\documentclass[draft]{agujournal2019}
\usepackage{url} 
\usepackage{lineno}
\usepackage[inline]{trackchanges} 
\usepackage{soul}
\usepackage{amsmath}
\journalname{JGR: Space Physics}

\begin{document}

%
%


\title{The Diffuse Auroral Eraser}

%
%




\authors{R. N. Troyer \affil{1}, A. N. Jaynes \affil{1}, S. L. Jones \affil{2}, D. J. Knudsen \affil{3},\newline T. S. Trondsen \affil{4}}


\affiliation{1}{Department of Physics and Astronomy, University of Iowa}
\affiliation{2}{NASA Goddard Space Flight Center}
\affiliation{3}{Department of Physics and Astronomy, University of Calgary}
\affiliation{4}{Keo Scientific Ltd.}




\correspondingauthor{Riley Troyer}{riley-troyer@uiowa.edu}




\begin{keypoints}
\item We identify an interesting and unusual feature, called a diffuse auroral eraser, which occurred during low magnetic activity.
\item An auroral eraser starts as diffuse aurora that brightens, dims to lower than its initial level, then recovers to its initial brightness.
\item We found the average recovery time to be 20 seconds with a 13 second standard deviation.
\end{keypoints}

%
%

%
%


\begin{abstract}
The source of diffuse aurora has been widely studied and linked to electron cyclotron harmonic (ECH) and upper-band chorus (UBC) waves. It is known that these waves scatter 100s of eV to 10s of keV electrons from the plasma sheet, but the relative contribution of each wave type is still an open question. In this paper, we report on an interesting and unusual auroral feature observed on March 15, 2002. We believe that these observations could help further our understanding of waves associated with diffuse aurora. This diffuse auroral feature is characterized by four phases: (1) the initial phase exhibiting regular diffuse aurora, (2) the brightening phase, where an east-west auroral stripe rapidly brightens, (3) the eraser phase, where the stripe dims to below its initial state, and (4) the recovery phase, where the diffuse aurora returns to its original brightness. Using a superposed epoch analysis of 22 events, we calculate the average recovery phase time to be 20 seconds, although this varies widely between events. We hypothesize that the process responsible for these auroral eraser events could be the result of chorus waves modulating diffuse aurora.   
\end{abstract}

\section*{Plain Language Summary}
\noindent Aurora are caused by electrons from within the magnetic bubble that surrounds Earth, called the magnetosphere. Sometimes these electrons are deposited into our upper atmosphere, producing light. This process can also transfer large amounts of energy from the magnetosphere to the atmosphere, thus potentially affecting the climate. Pictures of the aurora usually depict discrete green curtains, but this is not the only type of aurora. Diffuse aurora is another, which look like a faint glow over large portions of the sky. Diffuse aurora are extremely common, but not well understood. We have found an unusual process within diffuse aurora that could improve our understanding. We call it a diffuse auroral eraser. We found these events in a movie taken the night of March 15, 2002 in Churchill, MB, Canada. They appear as a section of diffuse aurora that rapidly brightens, then disappears and also erases the background aurora. Then, over the course of several tens of seconds, the diffuse aurora recovers to its original brightness. We calculated the average recovery time by overlaying plots of brightness from each of the 22 events that we found. This average time was 20 seconds, although it varied widely between individual events.

%
%

%


%
%
%
%

\section{Introduction}
Originally referred to as mantle aurora, diffuse aurora appears as a faint glow, just visible to the naked eye, and spread across a large portion of the sky \cite{lui1973}. The location of this aurora occurs equatorward of the discrete auroral oval and usually peaks in activity and brightness after magnetic midnight. This typically corresponds to between $60^\circ$ and $75^\circ$ corrected geomagnetic latitude, depending on the solar cycle and geomagnetic activity \cite{Sandford1968, Feldstein1985}. Work by \citeA{Sandford1968} showed, by integrating the 3914\AA{} emission over the auroral zone, that during solar maximum, diffuse aurora accounts for 80\% of auroral emission, however during solar minimum that was reduced to only 50\%. While discrete aurora are much brighter and better known, diffuse aurora play an important role in the magnetosphere - ionosphere (MI) system because they are so common and thus are one of the largest sources of energy transfer between the two regions of geospace \cite{newell2009}. The glow of diffuse aurora is the result of 100s of eV to 10s of keV electron precipitation from the plasma sheet \cite{Meng1979}. Many studies have identified wave-particle interactions in the plasma sheet as the primary way that this precipitation occurs. The waves responsible are electron cyclotron harmonic (ECH) and upper-band chorus (UBC) waves \cite <e.g.,>[and others]{Meredith2009, Thorne2010, Ni2016}. ECH waves are electrostatic perturbations whose frequencies fall between harmonics of the electron gyrofrequency $(f_\text{ce})$ \cite <e.g.,>[and others]{Kennel1970, Fredricks1973, Shaw1975, Gurnett1979}. UBC waves are electromagnetic and a subset of whistler mode chorus waves with frequencies between $0.5 f_\text{ce} < f < f_\text{ce}$ \cite <e.g.,>[and others]{Tsurutani1974, Burtis1976}. They differ from lower-band chorus (LBC) waves, which cover the frequency range $0.1f_\text{ce} < f < 0.5f_\text{ce}$. ECH and UBC waves also scatter slightly different energy electrons. ECH waves are most efficient between a few hundred eV to a few keV while chorus are most efficient below a few hundred eV \cite{Horne2003}. While most studies have linked ECH and UBC waves as the primary source of diffuse aurora, there is also a case to be made for whistler mode hiss waves, which have frequencies below $0.1f_\text{ce}$. These waves resonate best with electrons of energies above a few keV \cite{Horne2003}.   

Despite their usual static appearance, diffuse aurora often occur alongside the more dynamic pulsating aurora \cite{Davis1978}. Typical pulsating aurora are characterized by widely varying diffuse-like patches that blink on and off with periods between 2 to 20 seconds \cite <e.g.,>[and others]{Davis1978, Lessard2012} and can be widespread and long-lasting \cite{Jones2013}. While they frequently happen during diffuse aurora, studies have shown that pulsating aurora originate from a different wave-particle interaction. This interaction happens in the outer radiation belt between LBC waves and a few to 100s of keV electrons \cite <e.g.,>[]{Nishimura2010, Nishimura2011, Jaynes2013, Kasahara2018}. Another, less studied, phenomenon associated with diffuse aurora is black aurora. These are defined regions within an auroral patch that have no emissions \cite{Davis1978, Trondsen1997, Fritz2015}. The exact cause of these aurora is still unknown. 

Both typical pulsating aurora and black aurora appeared on the night of our data, although our data did not capture the typical pulsating aurora. However, our focus was on an unusual phenomenon, similar to \citeA{Dahlgren2017}, who used a high frame-rate, narrow field-of-view camera to study the rise and fall times of pulsating auroral patches. They reported several examples of a dip in intensity to below the background ``off" level, which then recovered to the background over several seconds. The events we report on are characterized by a diffuse arc, which appears, then rapidly disappears, blacking out the diffuse aurora in the region. However, these events differ from \citeA{Dahlgren2017} in that they tend to be isolated instead of periodic in time, although several events we observed did repeat for several periods. It is not clear that they are part of the same phenomenon. We quantify and report here the statistical distribution of recovery times of these features, which we call ``eraser" events. The events appear in images taken during a 2 hour period the night of March 15, 2002, during a campaign in Churchill, MB, Canada. While the campaign lasted for several days, this was the only night that these events appeared. In addition, during that night we only observed these events during this 2 hour window. This auroral feature is worth investigating since it could lead to a better understanding of diffuse aurora and the associated waves. What process in the equatorial magnetosphere can turn off diffuse auroral emissions in localized regions?

\section{Data}
\begin{figure}
 \noindent\includegraphics[width=\textwidth]{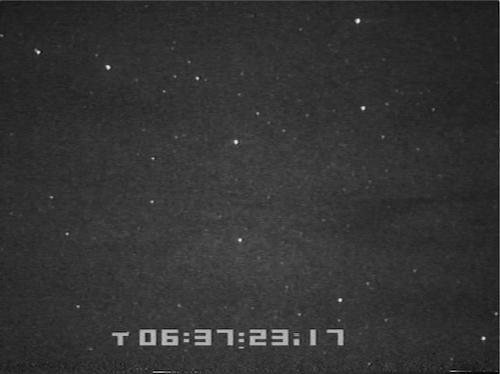}
\caption{An unprocessed image from the original data. The first several seconds of the video include a UT timestamp. Ursa Major appears at the top of the image.}
\label{fig:example_frame}
\end{figure}

The images used in this analysis were taken from an intensified, narrow field-of-view (FOV) CCD-based TV camera known as the Portable Auroral Imager (PAI) \cite{Trondsen1997}. They were taken at 30 frames-per-second (fps) and span from approximately 6:40 to 8:40 universal time (UT) or 0:06 to 2:06 magnetic local time (MLT) on March 15, 2002. The PAI was mounted to a tripod, which was on the ground at a location near Churchill, Manitoba. This corresponds to $69.28^\circ$ latitude and $331.22^\circ$ longitude in AACGM coordinates. It was pointed manually during the nightly observations and equipped with a 25 mm lens in addition to a Wratten 89B IR filter with a cutoff wavelength of 650 nm. \citeA{Trondsen1998} describes the PAI used to collect the data in more detail. This setup resulted in a $30.9^\circ$ by $23.2^\circ$ FOV. Given an image size of 640 by 480 pixels, the single pixel resolution is 88 m by 88 m at an altitude of 105 km \cite{Trondsen1998}. Using a section of Ursa Major that was visible in the images, we were able to estimate that the PAI was facing south of zenith, spanning elevation angles between $66.8^\circ$ and $90^\circ$. Figure \ref{fig:example_frame} is an example of one of the raw images. This is a white-light image and the auroral features are faint, but visible to the naked eye, however we have no additional spectral information. To reduce data size and improve image quality, we averaged every 10 frames to produce a 3 fps video, which we then analyzed. Our analysis of this data is a continuation of preliminary work by \citeA{Jaynes2013_2}. It is also worth noting that this data was taken during a targeted and short-lived campaign, where data was collected for only a select period each night. 

\section{Case Study}
\begin{figure}
 \noindent\includegraphics[width=\textwidth]{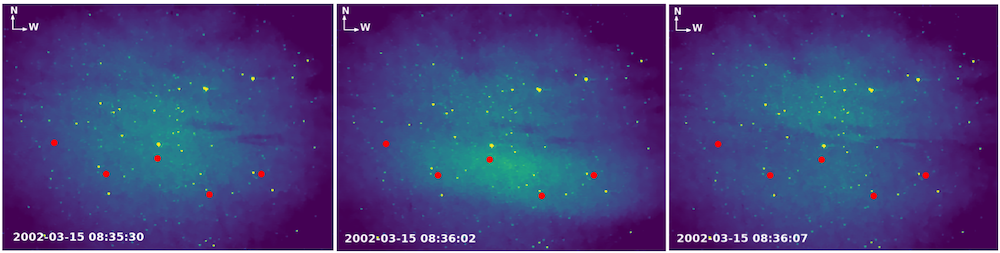}
\caption{Images across a characteristic eraser event. From left to right the images are before the event, during the event brightening, and during the event eraser. The red dots indicate the 5 pixels we used to represent the event.}
\label{fig:example_event_pics}
\end{figure}

To better describe this phenomenon we first investigate a single representative event (number 28 from the list of 32 events as identified in Figure \ref{fig:keogram}). Figure \ref{fig:example_event_pics} shows three images taken at different times during this event. We changed the contrast and color map of the images to see the event better. In the first image the aurora begins as a diffuse background. In the second image a diffuse auroral arc appears in the form of a brighter stripe in the lower third of the frame. By the third image, the stripe has disappeared and the background aurora is darker than the first image, as if someone has taken an eraser to it. Small black auroral forms can also be seen in each frame, although we do not explore those features in this paper.

\begin{figure}
 \noindent\includegraphics[width=\textwidth]{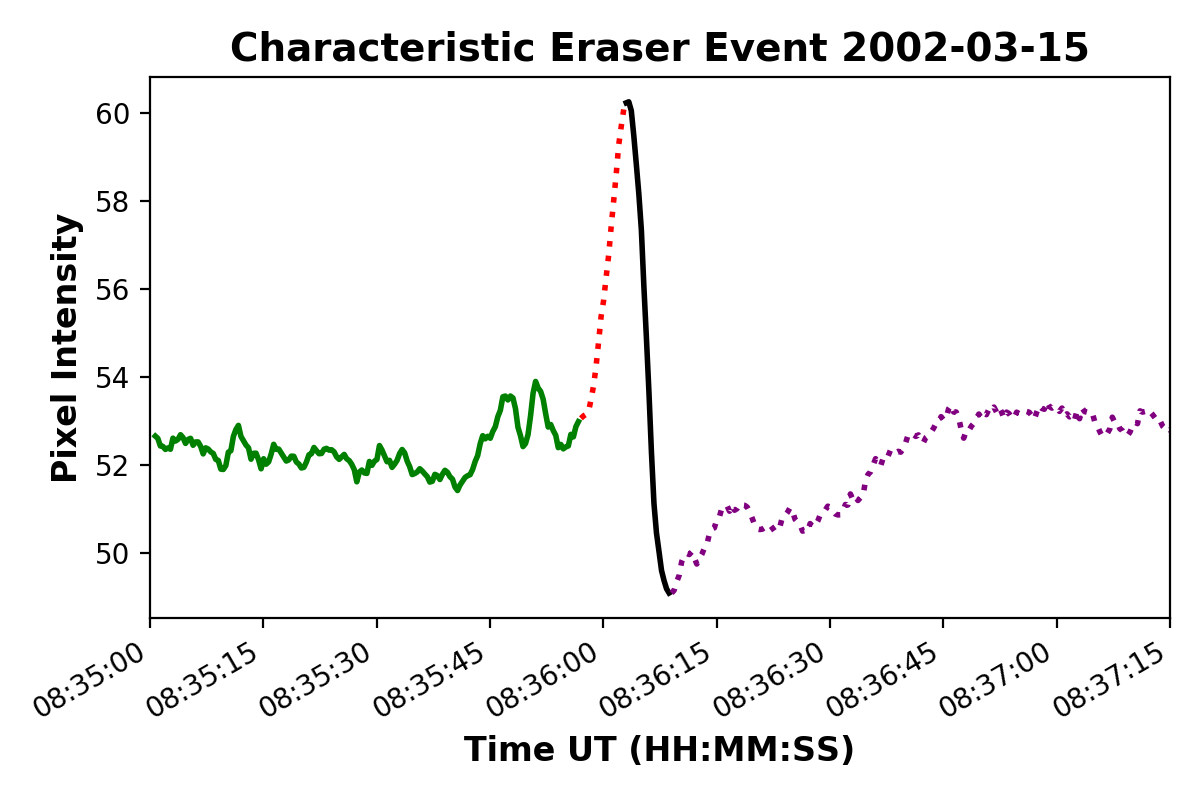}
\caption{The averaged intensity of a characteristic eraser event. The event is color coded by phase: initial (solid green), brightening (dotted red), eraser (solid black), recovery (dotted purple).}
\label{fig:example_event}
\end{figure}

To represent the event, we picked 5 pixels from across the image frame. These are represented by the red dots in Figure \ref{fig:example_event_pics}. By taking the median of a 5x5 pixel block centered on each of those, and then the average of the 5 block values, we were able to estimate the brightness of the event at each time step. Figure \ref{fig:example_event} shows this brightness plotted versus time after applying a 1.5 second smoothing window. In addition, Figure  \ref{fig:example_event} is color coded to indicate what we refer to as the 4 phases of an auroral eraser event. The initial phase (solid green) is the period before the event with a uniform diffuse aurora background. The brightening phase (dotted red) comes next and is characterized by an east-west stripe of aurora that rapidly brightens. Shortly after, the brighter section disappears in the eraser phase (solid black), taking the diffuse background with it. Finally, in the recovery phase (dotted purple) the pixels return to their original brightness over several tens of seconds. For this event, the recovery time is 32 seconds. The details of how this is calculated are in the next section.

\section{Analysis}
\begin{figure}
 \noindent\includegraphics[width=\textwidth]{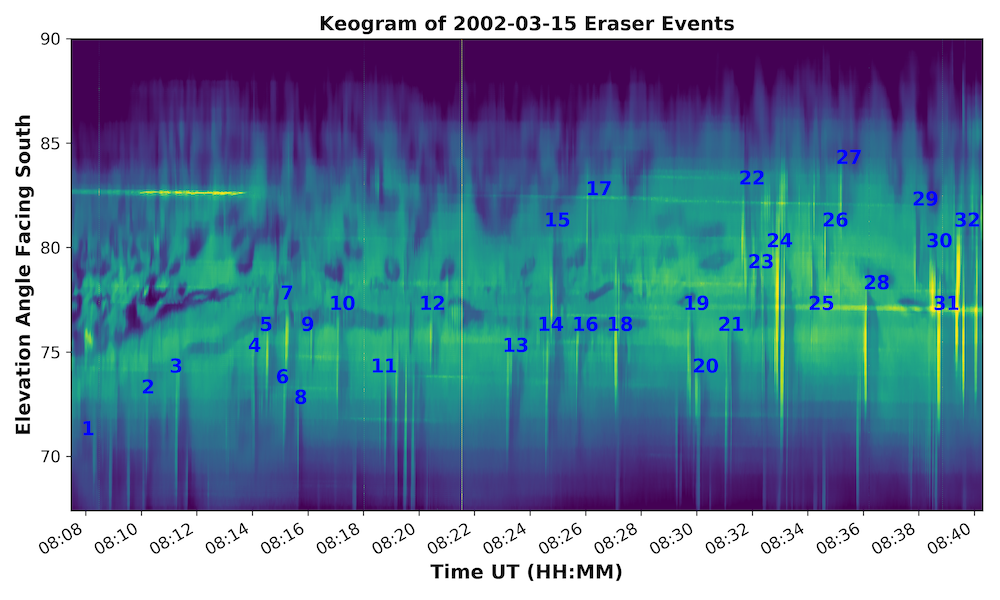}
\caption{A keogram containing all of the auroral eraser events we identified. We increased the contrast and used the Viridis color map to see the events better.}
\label{fig:keogram}
\end{figure}

We found that the best way to identify an auroral eraser event was in a keogram. This is a type of figure that is often used to visualize the evolution of aurora. A keogram is constructed by taking a north-south line of pixels from each image and aligning all of them chronologically. For each of our keogram columns we calculated a median column from the 21 center pixel columns of the corresponding image.  We did this to highlight the diffuse auroral features over the much brighter stars, which appear as bright horizontal lines throughout the keogram. Initially, we split the 2 hours of data into approximately 20 minute sections, and made a keogram for each. We found 32 auroral eraser events in the last 2 sections and no events preceding this. The keogram shown in Figure \ref{fig:keogram} spans a time range that contains all 32 events. These are labeled in the figure and are identifiable by a bright vertical strip, followed by a darker section. An observant reader might also notice darker patches scattered across the keogram in addition to the auroral erasers. These are black aurora, which we mention, but did not study further. Finally, the bright horizontal streaks are stars that happened to fall within the center columns of the base images.

\begin{figure}
 \noindent\includegraphics[width=\textwidth]{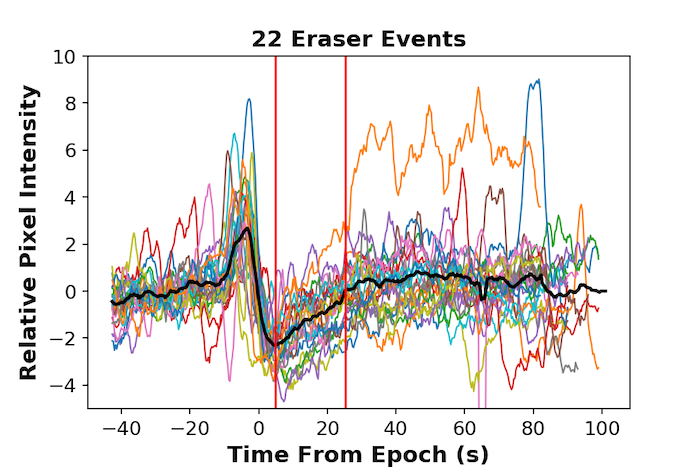}
\caption{Epoch analysis for the 22 eraser events with a full recovery. The red lines indicate the start and stop of the average recovery period, which is 20 seconds. Some of the time series show periodic behavior.}
\label{fig:epoch}
\end{figure}

\begin{figure}
 \noindent\includegraphics[width=\textwidth]{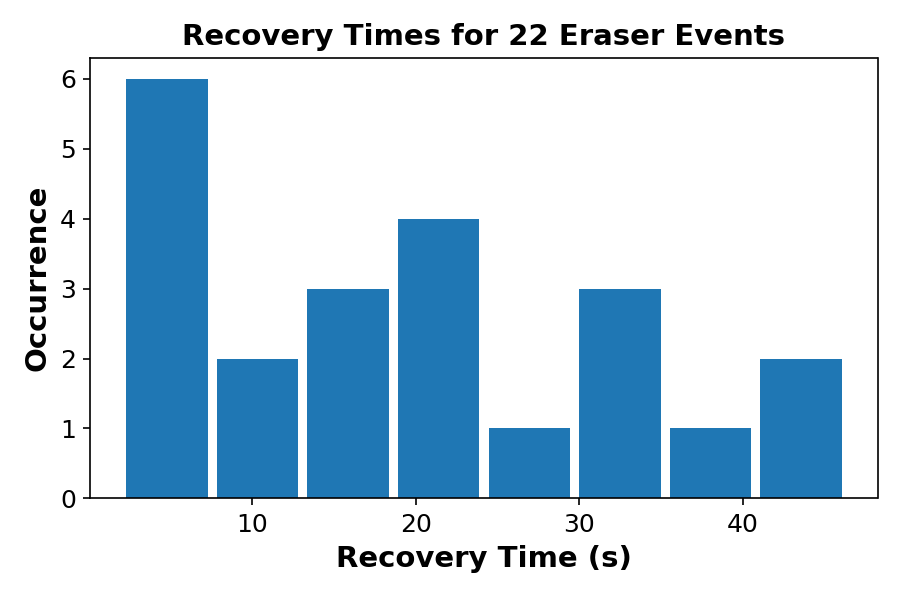}
\caption{Histogram of the recovery times for the 22 eraser events with a full recovery. The recovery time is calculated as the time between the end of the eraser phase to when the brightness returns to the average of 43 seconds to 10 seconds before the brightening phase.}
\label{fig:hist}
\end{figure}

Using the keogram to identify all of the events, we then created an intensity versus time plot, like Figure \ref{fig:example_event}, for each auroral eraser. Looking at these plots, it was clear that some events didn't have a full recovery phase, often being disrupted by a second event (see Figure \ref{fig:double_event} as an example). From the 32 events, we were able to visually identify 22 that returned to the original diffuse auroral brightness without being interrupted by a second event. Since we were interested in characterizing the recovery time, we limited our data set to these events. We then performed a superposed epoch analysis, shown in Figure \ref{fig:epoch}. We set an epoch time halfway between the peak of the brightening phase and the trough of the eraser phase. We also normalized the pixel intensity by setting the average of the initial phase to zero. The time range associated with this was 100 to 30 frames or 43 to 10 seconds before the epoch. We could then determine the time from trough of the eraser phase to when the brightness returned to zero, we called this the recovery time. The average recovery time from this analysis was 20 seconds. However, the standard deviation was 13.17 seconds, highlighting that individual events can vary dramatically. This can be further seen in Figure \ref{fig:hist}, which is a histogram of the recovery times from the 22 events. The large peak for times less than 10 seconds is a result of events whose initial phases slowly increase in brightness instead of remaining constant. This causes the baseline average to be lower than other events.  

Using the keogram we measured the approximate north-south extent of each eraser event, visually identifying the event edges. Widths range between 5 and 13 kilometers, with an average of 4.6 kilometers. This scale often shrinks as the eraser event fills in from the outsides during the recovery phase.

\begin{figure}
 \noindent\includegraphics[width=\textwidth]{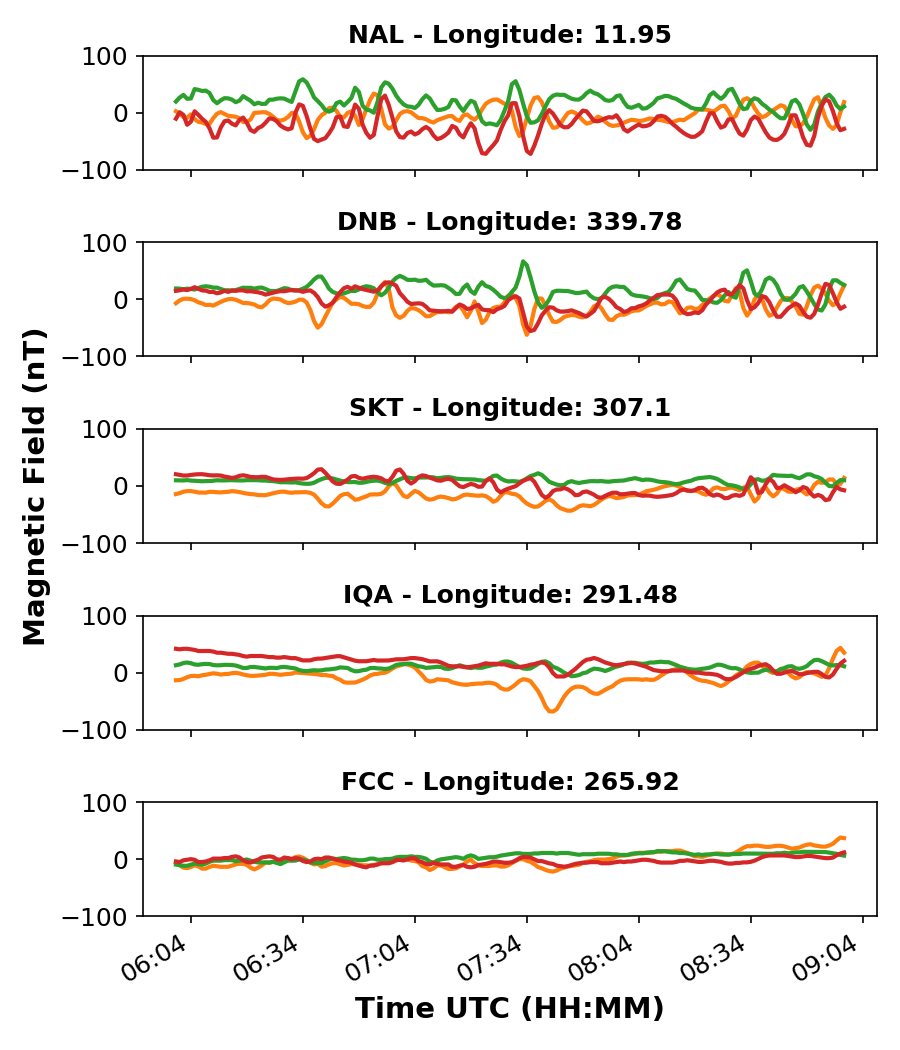}
\caption{Five ground-based magnetometers during the night of March 15, 2002. The plots are arranged by eastward geographic longitude from bottom to top. The bottom three are located between Churchill and the west coast of Greenland. The top two are located on the east coast of Greenland and on Svalbard.}
\label{fig:magnetometers}
\end{figure}

We were also interested in the magnetic conditions on March 15, 2002. To learn about these we looked at several different sources. The solar wind speed as extracted from NASA/GSFC's OMNI data set through OMNIWeb, was between 340 and 360 km/s. OMNI is a database of combined solar wind parameters from multiple spacecraft including WIND and ACE. $B_z$ ranged between 0 and 6 nT, the $\text{K}_\text{P}$ index was $< 2$, and the Dst index was $>-20 \, \text{nT}$ \cite{King2005}. In all, the solar wind and magnetic conditions were unremarkable and indicated quiet activity. The story was the same for ground-based magnetometers. SuperMAG is a database of over 300 magnetometers, each of which measures magnetic fields in 3 directions \cite{Gjerloev2009, Gjerloev2012}. We looked at data from several stations during the night of March 15, 2002. Data from Churchill (FCC) and surrounding magnetometers (IQA, SKT) showed no major perturbations in any of the field directions. Magnetometers on the north-east coast of Greenland and on Svalbard (DNB, NAL) did show some activity in the form of small wave-like fluctuations. These are plotted in Figure \ref{fig:magnetometers}. Unfortunately, the highest fidelity option was 1 minute, so we were unable to see any higher frequency modulations. Finally, the Auroral Electrojet (AE) index, which is derived from the horizontal component of select magnetometers around the globe, indicated low magnetic activity with a value of $< 90$ nT.     

\section{Discussion}
Other types of events showing atypical fluctuations in brightness of diffuse aurora have been published before \cite{Yamamoto1983, Dahlgren2017}. \citeA{Dahlgren2017} mentions observing a dip in brightness to below the average, during the off phase of pulsating aurora. While the events discussed here are similar, we believe that the differences are significant enough to provide additional scientific insight.

Watching video of the eraser events, the question arises, what type of aurora are these associated with? The background appears diffuse-like, but the eraser events are more structured. One possibility is that they are an atypical form of pulsating aurora. While typical pulsating aurora appears as patches of aurora that turn on and off with periods up to 20 seconds, which repeat for many periods, studies have observed other, less common behaviors. For instance, \citeA{Johansen1966} classifies type A pulsating aurora as pulses with periods between 1 and 3 seconds that decrease in intensity with each pulse until they are no longer visible. This is usually no more than 5-8 pulses. \citeA{Royrvik1977} also specify that a pulsating aurora patch can undergo only a single pulse. In addition, dual-layer pulsating aurora could also be similar \cite{Royrvik1976, Trondsen1997}. \citeA{Trondsen1997} described it as a section of foreground diffuse aurora that turns off, revealing a background with structure in it. It is possible that these are auroral eraser events, but their camera FOV was significantly smaller ($7.4^\circ$ by $5.5^\circ$), so it is difficult to tell with certainty. Another possibility is that the eraser events are a type of structured diffuse aurora. Nightside structured diffuse aurora manifests as regular, parallel auroral stripes, brighter than the background according to \citeA{Sergienko2008}, who found that structured diffuse aurora is caused by precipitating electrons above 3-4 keV and linked these to whistler mode hiss waves. A future study to find more of these events and compare them with known types of aurora such as those in \citeA{Royrvik1977} and \citeA{Fukuda2016} would be interesting. There are also many additional features in this data set that could be researched further such as the black aurora and non-erasing diffuse-like forms.

Without further data, we do not wish to speculate on what type of aurora the eraser events are linked to. However, we would like to note that, to our knowledge, reports of auroral behavior such as this are rare. That is intriguing, as the global and local geomagnetic data we investigated appears to show ordinary conditions that likely occur frequently. In addition, the 32 events we observed happened within a 2 hour period suggesting that when the conditions are right numerous eraser events can occur. Are eraser events such as these common during times of low global magnetic indices and have other observations just overlooked them due to the narrow field-of-view and high sensitivity required to see them, or are they rare? A more detailed search would be necessary to answer this question.

\begin{figure}
 \noindent\includegraphics[width=\textwidth]{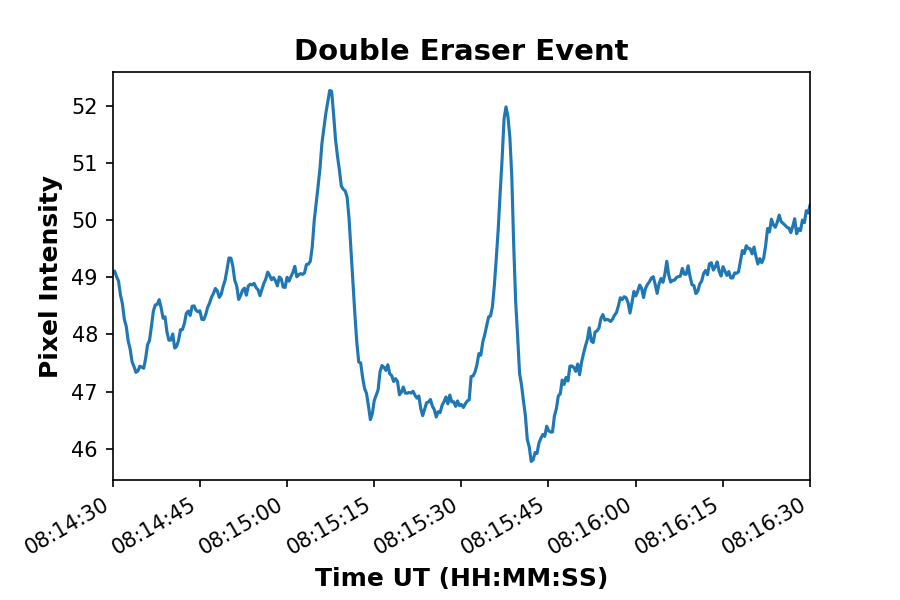}
\caption{Event 6 begins as a normal eraser event, but before the diffuse background is able to fully refill, a second event (7) appears to occur. After this the background refills as normal.}
\label{fig:double_event}
\end{figure}

Without concurrent in situ spacecraft observations, we can only speculate on possible drivers of these features. One possible cause could be modulated chorus waves interacting with a diffuse aurora population, possibly driven by ECH waves. As \citeA{ni2011_2} showed, the scattering efficiencies are different between ECH and chorus waves for different pitch angles. Chorus waves scatter more efficiently over a wider pitch angle distribution than ECH waves. If, during an instance of ECH-driven diffuse aurora, the particles drifted through a region of modulated chorus waves, the effect might be to momentarily increase the electron flux, causing the brightening we observe. If the chorus waves were strong enough they could deplete electrons in the narrower ECH pitch angle scattering distribution, causing the eraser. If this were true, we would expect a second peak if the particles drifted through another region of chorus waves before the diffuse aurora had completely recovered. We see this behavior in several cases as Figure \ref{fig:double_event} shows.

Another explanation could be the relaxation oscillator model proposed by \citeA{Davidson1986} to explain the periodic behavior of pulsating aurora. This model is driven by an interaction between the loss cone and scattering waves. When electrons are injected into a closed field line, waves begin to grow and scatter some of these into the loss cone. As the loss cone fills, wave growth halts. Without waves to scatter more electrons, the loss cone empties to a level lower than it started with. This might explain the erasing behavior of an eraser event. In this cycle, more particles are lost than would be required to reestablish equilibrium. For the process to continue for multiple cycles, the source of electrons needs to be strong enough to fill the loss cone back to equilibrium. 

These explanations are only speculative and we would need more detailed modeling and data to determine their feasibility.

\section{Summary}
\begin{itemize}
\item A diffuse auroral eraser event is characterized by an initial background of diffuse aurora, followed by the brightening of a more defined auroral stripe, which disappears and takes the background aurora with it. The background then refills back to the initial state at a slower rate. This recovery time can vary dramatically, but averages around 20 seconds. 
\item The  eraser events we observed occurred within diffuse aurora and during times of low magnetic activity. In addition, they all occurred within 40 minutes of each other.
\item From our knowledge there have been limited reports of events such as these, which raises the question: are these a common quite time phenomenon that has been overlooked or are they rare?
\end{itemize}

\acknowledgments
Imager data referenced in this paper is available from \citeA{Knudsen2002_data}.

Participation of DK in the March 2002 Churchill campaign was supported by the Natural Science and Engineering Council of Canada (NSERC). Participation of TT was supported by an NSERC grant held by LL Cogger. Participation of SJ was supported by Dartmouth College's Women in Science Project. DK, TT, and SJ thank the Churchill Northern Studies Centre for their support during the observing campaign. 

We acknowledge use of NASA/GSFC's Space Physics Data Facility's OMNIWeb service, and OMNI data (https://spdf.gsfc.nasa.gov/pub/data/omni/). The Dst and AE indices were provided by World Data Center C2, Kyoto University, Japan (http://wdc.kugi.kyoto-u.ac.jp/dstdir/). We gratefully acknowledge the SuperMAG collaborators (http://supermag.jhuapl.edu/info/?page=acknowledgement) for providing the magnetometer data.


%
%

\bibliography{iowa_bibtex}

%
%
%
%
%

\end{document}